# Divide-and-Conquer based Ensemble to Spot Emotions in Speech using MFCC and Random Forest

Abdul Malik Badshah, Jamil Ahmad, Mi Young Lee, Sung Wook Baik[*]

*College of Electronics and Information Engineering, Sejong University, Seoul, South Korea*
[*]Corresponding author email: sbaik@sejong.ac.kr

*Abstract*

*Besides spoken words, speech signals also carry information about speaker gender, age, and emotional state which can be used in a variety of speech analysis applications. In this paper, a divide and conquer strategy for ensemble classification has been proposed to recognize emotions in speech. Intrinsic hierarchy in emotions has been utilized to construct an emotions tree, which assisted in breaking down the emotion recognition task into smaller sub tasks. The proposed framework generates predictions in three phases. Firstly, emotions are detected in the input speech signal by classifying it as neutral or emotional. If the speech is classified as emotional, then in the second phase, it is further classified into positive and negative classes. Finally, individual positive or negative emotions are identified based on the outcomes of the previous stages. Several experiments have been performed on a widely used benchmark dataset. The proposed method was able to achieve improved recognition rates as compared to several other approaches.*

## 1. Introduction

Body poses, facial expressions, and speech are the most common ways to express emotions [1]. However, certain emotions like disgust and boredom cannot be identified from gestures or facial expressions but could be effectively recognized from speech tone due to the difference in energy, speaking rate, speech linguistic and semantic information [2]. Speech emotion recognition (SER) has attracted increasing attention in the past few decades due to the increasing use of emotions in affective computing and human computer interaction. It has played a tremendous role in changing the way we interact with computers over the last few years [3]. Affective computing techniques use emotions to interact in more natural ways. Similarly, automatic voice response (AVR) systems can become more adaptive and user-friendly if affect states of users could be identified during interactions. The performance of such systems highly depend on the ability to accurately recognize emotions. SER is a challenging task due to the inherent complexity in emotional expression.

Information regarding the emotions of a person can help in several speech analysis applications such as human computer interaction, humanoid robots, mobile communication, and call centers [4]. Emergency call centers all around the world deal with fake calls on a daily basis. It has become necessary to avoid wastage of precious resources in responding to fake calls. Utilizing information like age, gender, emotional state, environmental sounds, and speech transcript, the situational seriousness could be assessed effectively. If a caller report an abnormal situation by calling an emergency call center, the SER can be used to assess whether the person is under stress, fear or not which can increase the truth rate of person and can help the call centers in making an effective decision. The emotion detection work presented here is a portion of a large project for lie detection from speech in emergency calls.

This paper describes a divide-and-conquer (DC) approach to recognize emotions from speech using acoustic features. Three classifiers namely support vector machine (SVM) [5], Random Forest (RF) [6] and Decision Tree (DT) [7] were evaluated. Four models for each classifier were used at four stages to recognize the speech signals at each stage of the proposed method.

## 2. Related Work

Speech emotion recognition has been studied vigorously in the past decade. Several different strategies involving a variety of feature extraction methods, and classification schemes have been proposed. Some of the recent works in this area are being presented here.

Vogt and Andre [8] proposed a gender-dependent emotion recognition model for emotion detection from speech. They used 20 different features consisting of 17 MFCC, 2 energy coefficients, and 1 pitch value for gender-independent and 30 features consisting of 22 MFCC, 3 energy and 5 pitch value for male and female emotions to train Naïve Bayes classifiers. They evaluated their framework on two different datasets including Berlin and SmartKom. They achieved an improvement of 2% in overall recognition rates for Berlin dataset and 4% for SmartKom datasets, respectively. Lalitha et al. [9] developed a speaker emotion recognition system to recognize different emotions with a generalized feature set. They used and compared the result of two different classifiers including continuous hidden markov model (HMM) and support vector machine. Their proposed algorithm was able to achieve higher recognition rates for sad (83.33%), anger (91.66%), and fear (100%) on the validation set. However, emotions like boredom, disgust and happy were not recognized accurately and achieved 50%, 23.07 % and 26.31% accuracy rates, respectively. Shah et al. [10] conducted an experiment using Discrete Wavelet Transforms (DWTs) and Mel Frequency Cepstral Coefficients (MFCCs) for speaker independent automatic emotion recognition from speech. They achieved 68.5% recognition accuracy using DWT and 55% accuracy using MFCC. These accuracy rates were not satisfactory. Moreover, they only used four emotions namely neutral, happy, sad, and anger. Krishna et al. [11] also used MFCC and wavelet features. They used Gaussian Mixture Model (GMM) as classifier for emotion recognition. They concluded that using Sub-band based Cepstral Parameter (SBC) method provides higher recognition rates compared to MFCC. They achieved 70% recognition rate using SBC as compared to 51% using MFCC on their dataset.

Despite the efforts in constructing efficient SER systems using various features and classification frameworks, there is still significant room for improvement. Implementing SER for real world applications is a very challenging task which require careful design strategies [12].

## 3. Proposed Method

This section presents the overall architecture of the proposed framework, explains feature extraction method, and highlights key design aspects of the classification scheme.

### 3.1 Feature Extraction

Feature extraction is an important task in recognizing emotions from speech. The entire classification framework depends on the representational capability of the features being used. The proposed method uses acoustic features to recognize emotions. Mel Frequency Cepstral Coefficients (MFCCs) are robust to noise, possesses significant representational capability, and performs best with several speech analysis applications. Speech signals are divided into small frames of several milliseconds and 16 MFCC coefficients are extracted from each frame. These features are then used to train various classifiers for emotion modeling. MFCC features are well known and the readers are referred to [13, 14] for more details.

**3.2 Classification Framework**

The Proposed method use DC rule to detect and classify seven different types of emotions in speech. The emotion recognition system is divided into 3 stages. The first stage detects whether the sound signal contains any emotion or not (neutral speech). The second stage classify emotional signal into positive and negative emotions. The third stage further classify positive emotions into happy or surprise, and the negative emotions into angry, disgust, fear or sad shown in figure 1. This break down of emotions into several stages offer several advantages. Firstly, it simplifies the emotion recognition task by dividing in to several phases allowing us to effectively train efficient classifiers for each phase. Secondly, it reduces the amount of effort needed to detect emotions because if no emotions are spotted in the first stage, no further processing is required. Thirdly, heterogeneous classification schemes could be combined to form a high performance ensemble. Finally, the confidence of outcomes at the different phases can be used as evidence to compute the final results.

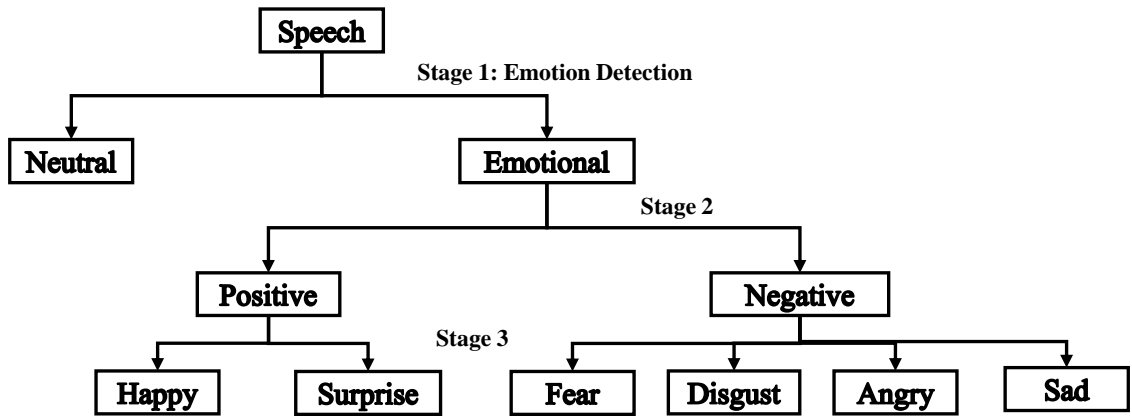

**Figure 1. Breakdown of emotional modeling for DC based classification**

For each stage in the classification framework, three different classifiers were evaluated. In phase 1, the objective is to detect emotions in speech. The training data was divided into neutral and emotional sets and the various classifiers were trained with it. Speech signals identified as neutral is not processed any further and the classification process ends immediately after phase 1. Furthermore, the classification schemes returned almost perfect recognition rate for neutral speech. The second phase attempts to further determine the nature of emotions being expressed by categorizing the emotional speech into positive or negative. Positive emotions include happy and surprise, whereas negative emotions include fear, anger, disgust, and sad. Finally, the positive or negative emotions into their respective base emotions. The DC based approach achieved better results as compared to using a single classifier for SER.

**4. Experiments and Results**

This section presents results of various experiments performed with the speech dataset for emotion recognition. It also highlights the major strengths and weaknesses observed during the evaluation process.

**4.1 Database Description**

Surrey Audio-Visual Expressed Emotion (SAVEE) Database [15] is used to evaluate the performance of the proposed method. The SAVEE database was recorded from four native English male speakers at the University of Surrey. It consist of 7 different type of emotions namely anger, disgust, fear, happiness, sadness, surprise and neutral. The sound files vary from 2 to 7 seconds in duration depending on the sentence uttered by the speaker. Total of 120 utterance per speaker were recorded.

**4.2 Experimental Setup**

The experiment was performed on a Desktop PC with Intel Core i7-3770 CPU @ 3.40GHz, 16.0 GB RAM and 64-bit Windows 7 Professional. The feature extraction and classification modules were implemented using C++. Experimental results analysis were performed in Matlab 2015a. Mainly two types of experiments were performed. The first experiment was carried out to compare the performance of DC based approach and single multiclass classifier based approach. The second experiment was to evaluate the proposed method through different classifier including SVM, Decision Tree and Random Forest. Results of these experiments are described in detail in the following sections.

**4.3 Stage-wise Classification Performance**

Stage wise classification performance is evaluated and discussed in this section. Fig. 2 shows the classification performance of DT, SVM, and RF. All these schemes achieved almost perfect recognition rates for neutral speech. However, recognition rates of emotional speech identification were relatively low with DT and SVM, achieving 81.11% and 80% accuracy, respectively. RF classifier achieved 92.2% accuracy for identifying emotional speech. Overall accuracy achieved by RF was the highest 96.11% at this stage.

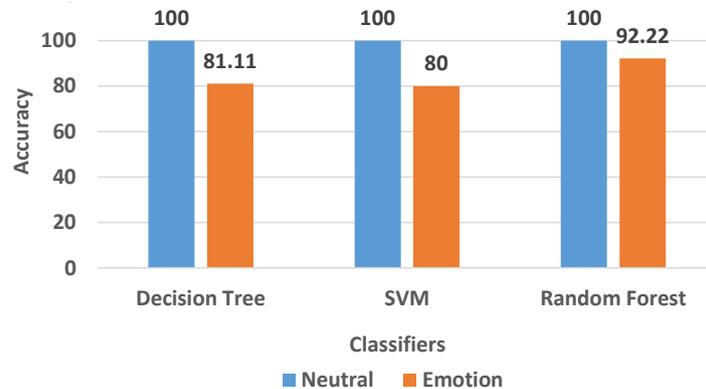

**Figure 2. Stage 1 classifier performance**

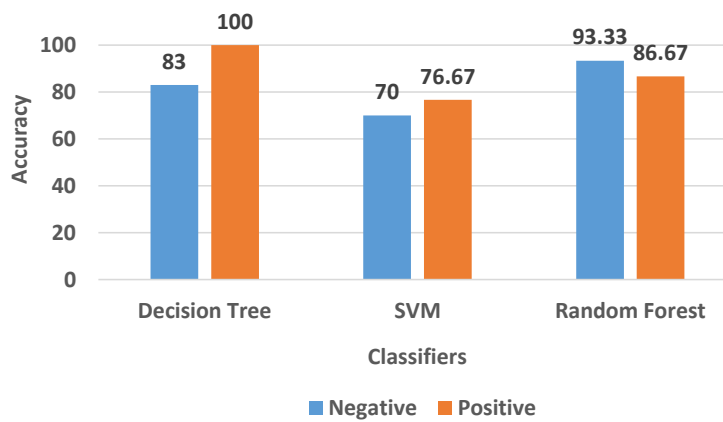

**Figure 3. Stage 2 classifier performance**

At Stage 2, classifiers are evaluated for identifying emotional speech obtained from previous phase as negative or positive. Accuracy rates are shown in Fig. 3. SVM Classifier has a lowest accuracy among the classifiers used at this stage for both negative and positive emotions, achieving accuracies around 70% and 76.67%. Decision Tree achieved better performance in identifying positive emotions. However, its performance with negative emotions is relatively low. RF achieved steady performance for both emotion classes achieving 90% overall accuracy at this stage.

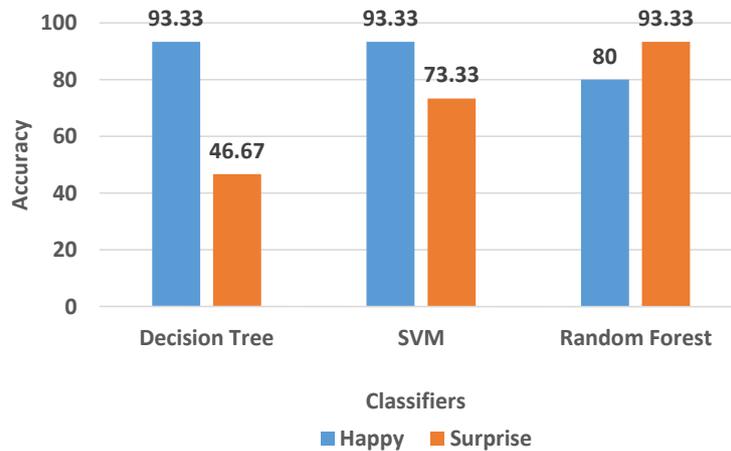

**Figure 4. Classification performance in stage 3 for identifying positive emotions**

Positive emotions from stage 2 are then identified as happy or surprise. Similarly, negative emotions are classified into angry, disgust, fear, and sad. Results in Fig. 4 reveal that RF classifier achieves higher emotion recognition accuracy as compared to DT and SVM. RF achieved 87% accuracy for identifying positive emotions. Though, DT and SVM offers higher accuracy in recognizing happy emotions but their performance in recognizing surprise emotion was relatively low, 73.33% accuracy rate for SVM and 46.67% for DT.

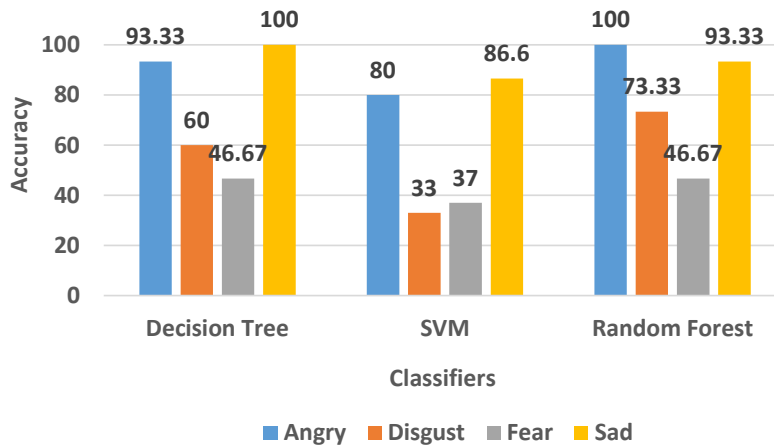

**Figure 5. Classification performance in stage 3 for identifying negative emotions**

A multi-class emotion recognition model was trained to classify negative emotions from stage 2 into angry, disgust, fear and sad. Results in Fig. 5 show that SVM classifier failed to classify fear emotions achieving only 37% accuracy, while its accuracy for angry, disgust and sad was 80%, 33% and 86.6%, respectively. DT achieved high accuracy in recognition of sad emotion, whereas its performance for angry, disgust and fear were 93%, 60% and 46.6%, respectively. In case of RF, angry, disgust, and sad emotions were successfully recognized with respective accuracies of 100%, 93.33%, 73.3%. However, the accuracy for fear emotion was relatively low at 46.67 %. Most of the test samples having fear emotions were miss-classified as disgust, as shown in Table 2. Comparatively, RF achieved high accuracy in classifying 4 different emotions with overall 78.3% recognition rate, Decision Tree achieved accuracy of 75%, and SVM was able to achieve accuracy of 60% only.

Overall classification results for using RF as a single classifier is provided in Table 1, and results of the proposed DC scheme is given in Table 2. By using RF as a single multi-class classifier for SER, accuracy rates were very low for disgust, happy, and sad emotions where each of these emotions were classified with less than 50% accuracy. Similarly, other emotions like angry, fear, and surprise achieved accuracy of about 66% each. The proposed DC based scheme was able to improve overall recognition accuracy for almost all emotions. From the performance comparisons of different classifiers, it can be easily concluded that RF is suited best among the classifiers for the proposed method.

**Table 1. Classification results for RF used as a single classifier**

| Emotion Class | Angry | Disgust | Fear | Happy | Neutral | Sad | Surprise |
|---|---|---|---|---|---|---|---|
| **Angry** | 66% | 7% | 14% | 0 | 0 | 0 | 13% |
| **Disgust** | 0 | 46.6% | 6.8% | 0 | 46.6% | 0 | 0 |
| **Fear** | 0 | 0 | 66.6% | 0 | 26.4% | 0 | 7% |
| **Happy** | 7% | 0 | 13% | 46.6% | 0 | 0 | 33.4 |
| **Neutral** | 0 | 0 | 0 | 0 | 100% | 0 | 0 |
| **Sad** | 0 | 0 | 0 | 0 | 54% | 46% | 0 |
| **Surprise** | 13 | 0 | 21% | 0 | 0 | 0 | 66% |

**Table 2. Overall classification results using DC based scheme**

| Emotion class | Angry | Disgust | Fear | Happy | Neutral | Sad | Surprise |
|---|---|---|---|---|---|---|---|
| **Angry** | 100% | 0 | 0 | 0 | 0 | 0 | 0 |
| **Disgust** | 0 | 73.33% | 13% | 0 | 0 | 13.7% | 0 |
| **Fear** | 13% | 27% | 47% | 0 | 0 | 13% | 0 |
| **Happy** | 0 | 0 | 0 | 80% | 0 | 0 | 20% |
| **Neutral** | 0 | 0 | 0 | 0 | 100% | 0 | 0 |
| **Sad** | 0 | 0 | 6.7% | 0 | 0 | 93.3% | 0 |
| **Surprise** | 0 | 0 | 0 | 6.7% | 0 | 0 | 93.3% |

### 4.4 Comparison with other SER approaches

Performance of the proposed scheme is compared with two similar approaches in Table 3. Results suggest that the proposed classification approach outperforms the other methods in identifying anger, disgust, neutral, and sad emotions. In case of fear, the proposed method performed better than [11], however, it was lower than the method described in [9]. Similarly, in case of happy emotion, our method achieved significantly better performance than [9] and slightly lower accuracy than [11]. Overall, the proposed method outperformed both methods [9] and [11] by 23% and 17%, respectively.

### 5. Conclusion

In this paper, we presented a divide and conquer based approach to detect and classify emotions from speech signals. MFCC features are extracted from the speech signals which are used to train classifiers at three different stages of the proposed scheme. At the first stage, generic emotions are detected in speech by using a simple binary classifier which separates neutral speech from the emotional speech. At the second stage, emotions are further identified as positive or negative. Finally, individual emotions are identified at the last stage of the classification scheme. Three different classification schemes including DT, SVM, and RF were evaluated at each of these three stages. From the evaluations, it is concluded that the proposed DC based emotions recognition scheme achieves high accuracy with RF classifier at each stage.

In future, we plan to incorporate Dempster-Shafer theory of fusion to combine the decisions of intermediate classifiers before making a final decision. We hope that this fusion will help to improve accuracy even further.

Table 3. Performance comparison with other relevant approaches

| Emotions | Emotion Recognition Accuracy | | |
|---|---|---|---|
| | **Proposed Method** | Method [9] | Method [11] |
| Anger | **100** | 91 | 90 |
| Boredom | - | 50 | - |
| Disgust | **73** | 23.7 | 70 |
| Fear | 47 | 100 | 20 |
| Happy | 80 | 26.31 | 90 |
| Neutral | **100** | 75 | 60 |
| Sad | **93.3** | 83.33 | 90 |
| Surprise | **93.3** | - | - |
| Overall (excluding Boredom and Surprise) | **82.21** | 66.5 | 70.0 |

**Acknowledgements**

This work was supported by the ICT R&D program of MSIP/IITP. (No. R0126-15-1119, Development of a solution for situation-awareness based on the analysis of speech and environmental sounds).

**Authors:**

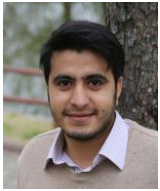

**Abdul Malik Badshah**

received his BCS degree in Computer Science from the Islamia College, Peshawar, Pakistan in 2013. Currently, he is pursuing Master degree in digital contents from Sejong University, Seoul, Korea. His research interests include speech analysis, machine learning, and applications of machine learning.

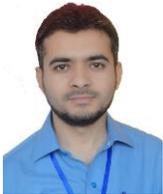

**Jamil Ahmad**

received his BCS degree in Computer Science from the University of Peshawar, Pakistan in 2008. He received his Master's degree in Computer Science with specialization in image processing from Islamia College, Peshawar, Pakistan. Currently, he is pursuing PhD degree in digital contents from Sejong University, Seoul, Korea. His research interests include image analysis, machine intelligence, semantic image representation and content based multimedia retrieval.

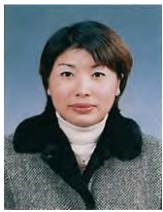

**Mi Young Lee**

is a research professor at Sejong University. She received her PhD degree in Image and Information Engineering at Pusan National University. Her research interests include Interactive Contents, UI, UX and developing Digital Contents. She has a MS in the Department of Image and Information Engineering from Pusan National University.

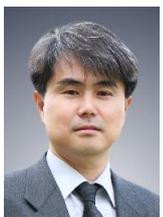

**Sung Wook Baik**

is a Professor in the Department of Digital Contents at Sejong University. He received the B.S. degree in computer science from Seoul National University, Seoul, Korea, in 1987, the M.S. degree in computer science from Northern Illinois University, Dekalb, in 1992, and the Ph.D. degree in information technology engineering from George Mason University, Fairfax, VA, in 1999. In 2002, he joined the faculty of the School of Electronics and Information Engineering, Sejong University, Seoul, Korea, where he is currently a Full Professor and Dean of Digital Contents. His research interests include computer vision, multimedia, pattern recognition, machine learning, data mining, virtual reality, and computer games.